Title: see http://summa.physik.hu-berlin.de/papers/TorstenAsselmeyer
-------------------------------------------------------------------------------
\documentstyle[12pt]{article}
\begin{document}
\title{Fractional Quantum Hall Effect, Composite Fermions and Exotic
  Spinors} \author{Torsten Asselmeyer$^{(1)}$ and Georg He\ss$^{(2)}$
  \\ $^{(1)}$ Institut f\"ur Physik, Humboldt Universit\"at zu Berlin
  \thanks{Invalidenstra\ss e 110, D-10099 Berlin, Germany, email:
  torsten@summa.physik.hu-berlin.de}\\ $^{(2)}$ Mathematisches
  Institut, Ludwig-Maximilians-Universit\"at M\"unchen
  \thanks{Theresienstr. 39, D - 80333 M\"unchen, Germany, email:
  hess@rz.mathematik.uni-muenchen.de}}
\date{June 14, 1995}
\maketitle
\begin{abstract}
\noindent
A topological interpretation of the electron--electron interaction
under FQHE conditions gives rise to the existence of some kind of
composite fermions.
The interaction term is similar to the Coulomb interaction and
produces punctures in the configuration space which change its topology.
Now the relevant configuration space is not simply connected and as a
consequence,
 ``new'' fermions, so called ``exotic spinors'', appear.
 We argue that composite fermions may be interpreted as exotic spinors.
The denominator of the filling factor is calculated to be an odd number.
We point out further possible relevance of exotic spinors.
\end{abstract}
PACS-numbers: 02.40.Ma, 02.40.Re, 02.40Vh, 73.40.Hm\\[0.3cm]
submitted to Journal of Physics A \\[0.4cm]
\noindent The quantum Hall effect (QHE) \cite{Kl} as a macroscopic
quantum phenomenon which appears in quasi--two dimensional electron
systems in a strong magnetic field and at very low temperatures shows
a certain universality; since the effect realizes an optimum between
high mobility and disorder , it is independent of the special material
and the geometry of the sample. Whereas the integer quantum Hall
effect (IQHE) leads to the quantization of the Hall conductivity with
integer numbers the fractional quantum Hall effect (FQHE) is
characterized by the fractional filling factor $\nu$ which is the
ratio of the number of electrons to the number of flux quanta (see
\cite{Sto} for theoretical models). In 1983 Laughlin \cite{Lau}
proposed a ground state wave function of the many body problem which
led to the correct filling factor $\nu = 1/2p+1$. This function
represents a new collective state of matter which is believed to arise
from condensation of the two--dimensional electrons to a quantum
liquid as a result of electron--electron interaction. Other fractions
are realised by an hierarchical scheme of parent and daughter states
(\cite{H1,H2}).  In Jain's composite fermion picture \cite{Jain} the
IQHE approach of noninteracting electrons and the FQHE approach with
electron correlations are conceptually unified, because he interprets
the FQHE as effect of noninteracting {composite fermions}. However,
the fundamental mechanism leading to the existence of composite
fermions is still unclear.\\ The purpose of this Letter is to present
a theoretical framework for the emergence of composite fermions as the
result of electron--electron interactions and spin--magnetic field
coupling. The fact that electrons are fermions is very important in
our picture. First of all, it provides a relation between massive
particles like electrons described by the Pauli Hamiltonian and
massless particles described by the Dirac Operator. In the case of
massless particles the interaction is encoded as a shift in the vector
potential. In fact, the main idea is that this interaction induces a
''nontrivial'' topology on the relevant configuration space. This
nontrivial topology is in turn responsible for the existence of
different types of fermions -- or, in more mathematical terms, of
spinors. \\ These so--called ''twisted spinors'' first appeared in
1978 in the context of quantum field theory and , in particular,
quantum gravity \cite{Ish1,Ish2}; in 1979, Petry \cite{Pet} introduced
the notion of ''exotic spinors'' in the context of
superconductivity.\\ In both cases, the essential feature is the
non-trivial topology of the configuration space under consideration.
Non-trivial topology means here that there exist circles in
configuration space that cannot be contracted to a point. In
mathematical terms those spaces are called {not} simply - connected.\\
In fact, there exist as many inequivalent spinors as there are
elements in the cohomology group $ H^1(M,{Z\!\!\!Z}_2) $ where $M \,$
is the configuration space; this particular group may be thought of as
somehow representing the inequivalent nontrivial circles and allowed
combinations of them. Perhaps the easiest way of viewing this problem
is the idea that on spaces with non-trivial topology there is an
additional degree of freedom for fermionic particles. Intuitively, one
may visualize this new degree of freedom in the following way:\\
Spinors may be twisted around the nontrivial circles, thus explaining
the notion of twisted spinors. Classically one is inclined to think of
these different spinors as describing different particles. On the
other hand it is also possible that in the quantum picture there
should be a sum over all possibilities. This is more in the spirit of
Avis and Isham \cite{AvIs} who proposed in the context of quantum
field theory a new partition function which in some respect is a sum
over all these different possibilities. \\ In 1979, Petry realized
that -- under some additional assumptions on the topology of $M \,$ --
it is possible to translate the different spinors into one chosen
spinor frame.  Essentially, the twisted spinors behave {after}
translation like normal spinors with the {only important difference}
that {the covariant derivative acting on these spinors has to be
  changed by an additional one-form} representing in some way the
non-trivial topology ( or , more precisely, an element in the above
mentioned cohomology group).\\ These translated twisted spinors are
also called exotic spinors; it is precisely the additional one-form
appearing in the covariant derivative {in a natural way as a
  consequence of topology} which makes exotic spinors so interesting
in the context of composite fermions.\\ In this letter we cannot give
a full derivation of the corresponding formulae but simply write down
the expression most important for our further arguments; for a
thorough and explicit derivation, unfortunately in a $(3 +
1)$--dimensional model, see \cite{Pet} and \cite{Ish1,Ish2}. The
covariant derivative $\nabla^{e} \,$ of exotic spinors $\psi^{e} \,$
in the direction of a vector field $Y \,$ is given by
\begin{eqnarray}\label{uff}
  \nabla^{e}_{Y} \psi^{e} = \nabla_{Y} \psi^{e} - \frac{1}{2} \, \big[
  \iota(Y) \, \, \lambda^{-1} \, \mbox{\rm d} \,\lambda \, \,\big]
  \,\psi^{e}
 \end{eqnarray}
 where $\nabla \,$ is the usual spin connection and $\iota \,$ denotes
 contraction of a one--form with the given vector field. Note the
 interesting factor $ \frac{1}{2} \, $ as a consequence of the
 mathematics involved. Here, the one--form $\lambda^{-1} \, \mbox{\rm
   d} \,\lambda \,$ with $\lambda :M \, \longrightarrow \, U(1) \,$
 which is closed but not exact is chosen in such a way that it
 represents the particular nontrivial element in the above mentioned
 cohomology group responsible for its existence. (On compact
 Riemannian manifolds this can be done in an essentially unique way
 \cite{AlEe}.) Therefore the integral over circles in the
 configuration space of the one--form $\frac{1}{2 \pi \, i} \,
 \lambda^{-1} \, \mbox{\rm d} \,\lambda \,$ takes only values $ 1 \,$
 or $ 0 \, $ depending on whether the circle can be deformed to the
 particular non--contractible circle or not. As an aside we remark
 that the existence of the additional one--form is related to
 mathematical considerations that resemble locally singular gauge
 transformations.  However, we point out that the above arguments are
 all well--defined in the theory of fibre bundles and are just the
 mathematical image of the nontrivial topology. We certainly do not
 want to advocate the use of singular gauge transformations as a - in
 general - well defined concept!\\ We consider the spin--magnetic
 field coupling as described by the Pauli Hamiltonian
\begin{eqnarray} \label{pauli}
  h_P=\left(\frac{1}{2M}\right)(-i\hbar\nabla - e \vec{A})^2 +
  \left(\frac{e\hbar}{2M}\right)\vec{\beta}\cdot \vec{B}
\end{eqnarray}
with $\vec{A}$ the vector potential, $\vec{B}=\nabla\times \vec{A}$
the magnetic field and $\vec{\beta}=(0,0,\pm 1)$ the spin direction.
The ground state is degenerate with respect to the number of flux
quanta \cite{SeSo}.  In the many particle picture we have to introduce
an interaction term. The configuration space of the 2d--electrons is
the plane $\mbox{\rm I\kern -.2em R}^2$ and with respect to the
experimental situation we impose periodic boundary conditions.  This
leads to the torus $T^2$ as configuration space and to the
asymmetrical gauge for the vector potential $\vec{A}$. Let $z_i,
z_j\in\mbox{\rm C}$ be the coordinates of two electrons given as
complex numbers.  Next we choose the interaction term to be given in
terms of complex numbers as the shift
\begin{eqnarray} \label{pot}
  A(z_i) \longrightarrow A(z_i)-\frac{\hbar}{e}\sum\limits_{j,\,
    i\not= j} \frac{1}{z_i-z_j}
\end{eqnarray}
in the vector potential and in the derivative, respectively. The same
term appears in the model of FQHE as a conformal field theory with
${\cal W}_{1+\infty}$--symmetry \cite{FV93}.\\ Due to the interaction
each electron sees the other electrons as punctures and hence the
configuration space will be topologically nontrivial. None of the
closed curves around a given puncture can be continuously deformed to
a point. We stress that nontriviality caused by punctures appears only
in two dimensions because a closed curve around a puncture in three
dimensions {can} be deformed to a point (using the third dimension for
the deformation).  First we consider the case of two particle
interaction given by the potential (\ref{pot}). Here we have to remove
two points from the configuration space.  However for further
topological considerations it is better to remove a little disk
instead of a point to get a compact configuration space $M$ again.
Therefore, two new generators of $H^1(M,{Z\!\!\!Z}_2)$ appear
corresponding to the number of removed disks or particles,
respectively. Since the translation procedure respects the group law
of $H^1(M,{Z\!\!\!Z}_2) \,$ we get classically three types of exotic
spinors with additional one--forms $\frac{1}{2} \, \lambda^{-1} \,
\mbox{\rm d} \,\lambda \,$ , $\frac{1}{2} \, \alpha^{-1} \, \mbox{\rm
  d} \,\alpha \,$ and $\frac{1}{2} \, \left( \lambda^{-1} \, \mbox{\rm
  d} \,\lambda \, + \, \alpha^{-1} \, \mbox{\rm d} \,\alpha \right)
\,$, respectively.  Due to (\ref{uff}) the derivative in (\ref{pauli})
is shifted for the different exotic spinors by the corresponding
one--form given above.  Following \cite{Pet} and \cite{Ish1} we
interprete this one--form as a global vector potential generated by
the nontrivial topology which does not contribute to the internal
magnetic field (because the one--form is always closed ).  We stress
that due to the nontrivial topology the theorem of Stokes does not
apply here in the usual sense.  Therefore we define the total flux
$\Phi$ through our system as the line integral of the relevant
connection one-form along the boundary of the configuration space.
\begin{eqnarray}
  \Phi = \oint\limits_{\partial M} A(z) dz = \oint\limits_{\partial
    D_1\cup \partial D_2} A(z)dz = \frac{h}{e}\cdot q \qquad
  q=1,2,\ldots
\end{eqnarray}
where $\partial D_1,\partial D_2$ denote the boundaries of the removed
disks. If we now assume that all different spinors (spin structures)
contribute equally we find:
\begin{eqnarray}
  \Phi\cdot\frac{e}{h} &=& \oint\limits_{\partial D_1\cup \partial
    D_2} \frac{e}{h}A(z)dz + \oint\limits_{\partial D_1\cup \partial
    D_2} (\frac{e}{h}A(z)dz - \frac{i}{4\pi}\lambda^{-1}\mbox{\rm
    d}\lambda - \frac{i}{4\pi}\alpha^{-1}\mbox{\rm d}\alpha) \nonumber
  \\ &+& \oint\limits_{\partial D_1} (\frac{e}{h}A(z)dz -
  \frac{i}{4\pi}\lambda^{-1}\mbox{\rm
    d}\lambda)+\oint\limits_{\partial D_2} (\frac{e}{h}A(z)dz -
  \frac{i}{4\pi}\alpha^{-1}\mbox{\rm d}\alpha)\nonumber \\ &=& q + (q
  - \frac{1}{2}-\frac{1}{2}) + (q - \frac{1}{2}) + (q -
  \frac{1}{2})=4q-2
\end{eqnarray}
where $\Phi\cdot\frac{e}{h}$ is the number of flux quanta. Although
this seems to be a very crude approximation we do find the same result
in the path integral formulation using the partition function of Avis
and Isham ( \cite{AvIs} , all weights equal to $1 \,$) once flux is
defined in the corresponding way. \\ This particular choice is also
supported by the work of Asch et al.  \cite{Asch} who show for the
special example of the torus that only the direct integral over all
non-equivalent connections on a suitable hermitian line bundle over
the torus with suitably chosen electromagnetic field strengths is
unitarily equivalent to the unique Bochner Laplacian on its universal
cover with the corresponding field strength.  This result can be
interpreted in analogy to our case dealing with twisted complex scalar
fields \cite{Ish1} instead of twisted spinors.\\ Since there are two
particles involved the inverse of the filling factor is given by
\begin{eqnarray}
\frac{1}{2}\Phi\cdot\frac{e}{h} = \frac{1}{\nu} = 2q-1
\end{eqnarray}
where $\nu$ is the filling factor. Following \cite{Ver} it seems
natural to assume that interaction is dominated by interaction of
pairs; then the total flux produced is just the flux $ N (4q - 2 ) \,$
for $ N \,$ {pairs} of fermions leading again to the factor $2q - 1
\,$ for $ \frac{1}{\nu} \,$.\\ In conclusion, a toy model of the
possible topological origin of some aspects of interaction was
presented giving rise to a kind of composite fermion picture and
providing an explanation of the odd denominator of the filling
factor.\\ The two--dimensionality is essential for the model because
only there the removal of points generates a nontrivial topology.
Characteristic for the particular model is that the number of
interacting particles is indeed equal to the number of holes in
configuration space which are in turn responsible for the existence of
exotic spinors. The other key ingredient is that we take all possible
spinor structures into account following the ideas of Avis, Isham
\cite{AvIs} and Petry \cite{Pet}.\\ The existence of exotic spinors
is, of course, not restricted to our model of the FQHE. They will -
under some additional topological assumptions - always exist if the
relevant configuration space is topologically nontrivial.  This is,
e.g., the case if small disks are removed from a 2 D configuration or
if small tubes are removed from a 3 D situation. (For interesting
experiments in the context of composite fermions, see e.g.
\cite{Flip} .)\\ But the same reasoning is also true for the full
torus ( with its topology $S^1 \, \times \, \mbox{\rm I\kern -.2em
  R}^2 \,$, see \cite{Pet} ). Indeed we cannot refrain from mentioning
that in this case exactly two different spinor structures
exist which would, in principle, give rise to half--integer flux quantization.
\section*{Acknowledgments}
The authors thank the Graduiertenkolleg ``Mathematik im Bereich
ihrer Wechselwirkung mit der Physik'' of the Ludwig Maximilian
University Munich , in particular Prof. Batt and Prof. Schottenloher,
for the possibility of a stay of T. A. at Munich where most of the
ideas took form.  G.H. is also indebted to Prof. Baum, Prof.
Friedrich, Prof. Petry and W. Posch for helpful discussions on exotic
spinors. T.A. thanks also Prof. Keiper for useful discussions.

\end{document}